\documentclass[11pt]{article}
\usepackage{amssymb,cite,epsf,graphics,graphicx}
\begin{document}

\title{Multicolored Temperley-Lieb lattice models.\\
 The ground state.\\}
\author{\textbf{A.Babichenko}$^{a,b}$\thanks{%
e-mail: babichen@wicc.weizmann.ac.il} and \textbf{D.Gepner}$^{a}$\\
\\
$^{a}$\textit{Department of Particle Physics,}\\
\textit{\ Weizmann Institute of Science, Rehovot 76100, Israel}\\
\\
$^{b}$\textit{Racah Institute of Physics,}\\
\textit{\ the Hebrew University, Jerusalem 91904, Israel}.\\}
\date{May 2006}
\maketitle

\begin{abstract}
Using the inversion relation method, we calculate the ground state
energy for the lattice integrable models, based on baxterization
of multicolored generalization of Temperley-Lieb algebras. The
simplest vertex and IRF models are analyzed and found to have a
mass gap.
\end{abstract}

\section{Introduction}

Looking for new solutions of Yang-Baxter equation is an
interesting and important problem in exactly solvable statistical
models. Some time ago set of new solutions of Yang-Baxter
equations were obtained \cite{DF} by "baxterization" of
multicolored generalization of Temperley-Lieb (TL) algebras -
Fuss-Catalan algebras -- discovered by Bisch and Jones \cite{BJ}.
These solutions are really new and are not equivalent to a fusion
of known models. In \cite{DF} these solutions were formulated in
terms of generators of this new algebra, and in principle the
Boltzmann weights may be interpreted both as a vertex model, or as
an interaction round the face (IRF) models defined on a tensor
product of admissibility graphs corresponding to different colors
of the TL algebra. Physical interpretation of these new integrable
models may be different. In addition to construction of new
integrable vertex and IRF models, an obvious possible realization
of these new solutions is multicolored generalization of dense
loop models \cite{DF}, polymers and other subjects of exactly
solvable statistical models (for a review see e.g. \cite{DF2}).
Another new integrable model of the same class was obtained as
integrable solution for Lorentz lattice gases \cite{M},\cite{MN}
and dimerised coupled spin-$1/2$ chain \cite{MN1}.

In spite of traditional form of the R-matrix, it turns out that
the usual algebraic Bethe ansatz  solution, at least in its naive
and straightforward version, reveals some technical problems. One
of the possible reasons for that might be a non trivial structure
of pseudo vacuum, which is hard to guess in a way effective for
realization of algebraic Bethe ansatz program. The alternative
method for investigation of physical properties of integrable
models, such as ground state energy, its free energy, spectrum and
correlation functions is the method of inversion relations. It
doesn't require explicit form of BA equations or transfer matrix
eigenvalues, valid in the thermodynamic limit. It was taken up by
Baxter (see, e.g. \cite{B1},\cite{B}) and further developed by
others (see e.g. \cite{P},\cite{K}). It exploits functional
inversion relations for the transfer matrix in the ground state,
and transfer matrix
 analytical properties, in order to calculate ground state energy
and low lying excitations spectrum \cite{K}. Assumptions on the
transfer matrix eigenvalue analytical properties usually should be
confirmed numerically. The advantage of this method is that it
doesn't require knowledge of Bethe ansatz equations and
eigenvalues of transfer matrix explicitly, but gives a possibility
to find basic physical quantities of the model quite easily.

Plan of the paper is the following. We start Sect.2 with
description of the Boltzmann weights of the models in terms of
multicolored TL generators and their algebraic relations. We
analyze possible physical regimes, where Boltzmann weights are
real and positive, and discuss possible vertex and IRF
interpretation of the models. Some interesting factorization
property of admissibility graph for IRF models is pointed out. In
Sect.3 we derive functional inversion relations for the transfer
matrices of two colored vertex model and for the partition
function of two colored IRF model. We solve them analytically as a
product of analytical non zero function, and a function with poles
or zeros. Excited states are discussed and found to lead to a mass
gap. In Sect.4 generalization of inversion relations for
multicolored models is discussed. We conclude by Sect.5 with a
discussion of the obtained results.

\section{Colored TL lattice models.}

In this section we recall the results of \cite{DF} and demonstrate
them for few examples of two, three and four colored models. For
details of
Fuss-Catalan algebras and their representations we refer a reader to \cite%
{BJ} and \cite{DF}. We also discuss vertex and IRF representations
of baxterized Fuss-Catalan models and find their possible physical
regimes.

\subsection{Solutions of Yang-Baxter equation with Fuss-Catalan symmetry.}

New class of TL like algebras found by Bisch and Jones \cite{BJ}
can be considered as a kind of quotient of tensor product of few
TL algebras. Generators of these algebras may be formulated in
terms of TL generators defined on a lattice. The relations on
generators depend on the parity of
the site $\ i$ where the operator acts. The main commutation relations \cite%
{BJ} are the following.%
\begin{equation}
U_{i}^{(m)}U_{i}^{(p)}=U_{i}^{(p)}U_{i}^{(m)}=\rho _{i}(\min
(m,p))U_{i}^{(\max(m,p))}  \label{u1}
\end{equation}%
where%
\begin{equation}
\rho _{i}(p)=\left\{
\begin{array}{c}
\alpha _{1}\alpha _{2}...\alpha _{p} \ \ i \ - \ even \\
\alpha _{k}\alpha _{k-1}...\alpha _{k+1-p} \ \ i \ - \ odd%
\end{array}%
\right.   \label{ro}
\end{equation}%
Here $k$ is the number of colors is, and in the standard strings
pictorial interpretation of the operators $U^{(m)}$ weight $\alpha
_{m}$ is attached to each loop of color $m$. Locality of operators
$U$ expressed as
\begin{equation}
U_{i}^{(m)}U_{j}^{(p)}=U_{j}^{(p)}U_{i}^{(m)}  \label{u2}
\end{equation}%
when $|i-j|>1$, or when $|i-j|=1$ and $m+p\leq k$. Non commuting
operators
satisfy cubic relations%
\begin{equation}
U_{i}^{(m)}U_{i\pm 1}^{(p)}U_{i}^{(q)}=\rho _{i}(k-p)\left\{
\begin{array}{c}
U_{i}^{(m)}U_{i\pm 1}^{(k-q)} \ \ m\geq q \\
U_{i\pm 1}^{(k-m)}U_{i}^{(q)} \ \ m\leq q%
\end{array}%
\right.   \label{u3}
\end{equation}%
These are basic relations of the Fuss-Catalan algebra
$FC_{k(N+1)}(\alpha _{1},\alpha _{2},...\alpha _{k})$ which
describes coils of $k(N+1)$ strings of $k$ colors. Boundary
conditions are fixed by requirement that both on the
top and on the bottom strings end up by the same pattern $%
a_{1}a_{2}...a_{k}a_{k}a_{k-1}...a_{2}a_{1}a_{1}...$. Pictorial
interpretation of the commutation relations written above one can
find in \cite{DF}. Lets note that not all of the relations written
above are independent (for details see \cite{BJ},\cite{DF}). There
are some consequences of these relations necessary, in particular,
if one wishes to check validity of Yang-Baxter equations for the
baxterization solutions, but we will not stop on these technical
details here.

In \cite{DF} baxterization of this algebra was successfully done.
Namely, all possible solutions of the Yang-Baxter equation
\begin{equation}
W_{i}(x)W_{i+1}(xy)W_{i}(y)=W_{i+1}(y)W_{i}(xy)W_{i+1}(x)
\label{yb}
\end{equation}%
were found in the form
\begin{equation}
W_{i}(x)=1_{i}+\sum_{m=1}^{k}a_{m}(x)U_{i}^{(m)}  \label{w}
\end{equation}%
satisfying
\begin{equation}
W_{i}(1)=1_{i}
\end{equation}%
in the class of rational polynomial functions $a_{m}(x)$. It was
done by substitution of (\ref{w}) into (\ref{yb}), which, using
the Fuss-Catalan algebra, leads to a complicated system of
functional equations on the coefficients $a_{m}(x)$. General
solution of this system of functional equations was found to be:
\begin{eqnarray}
a_{m}(x) &=&\frac{1}{\rho _{m}}x^{r_{1}+r_{2}+...+r_{m-1}}(x^{r_{m}}-1),\;%
\;m=1,2,...k-1.  \label{am} \\
a_{k}(x) &=&\frac{\rho _{1}}{\rho _{k-1}}x^{r_{1}+r_{2}+...+r_{k-2}+1}\frac{%
x^{r_{k-1}}-1}{\mu -x},  \label{ak}
\end{eqnarray}%
where $\rho _{m}\equiv \rho _{i}(m)$ are defined in (\ref{ro}),
$r_{m}=\{\pm 1\}$, and $r_{m}$ and other parameters entering the
last equations, are restricted and defined by the following set of
relations: $r_{1}=r_{k-1}=1$ and either
$r_{q-1}=r_{q}=r_{k-q}=r_{k-q+1}$, or
$r_{q-1}=-r_{q}=-r_{k-q}=r_{k-q+1}$,
 and
\begin{equation}
\mu ^{\left( r_{q-1}+r_{q}\right) /2} =\alpha _{q}^{2},  \label{mu} \\
\mu  =\alpha _{1}^{2}-1,  \label{al1} \\
\alpha _{q} =\alpha _{k+1-q}  \label{alal}
\end{equation}%
for any $q=2,...,k-1$. In \cite{DF} the special solution with $%
r_{1}=r_{2}=...=r_{k-1}=1$ was called fundamental. With this set
of $r_{q}$ one has $\mu =\alpha _{1}^{2}-1=\alpha
_{2}^{2}=...=\alpha _{k-1}^{2}$. These equations were solved in
the assumption that all $\alpha $'s are positive. We will relax
this condition and permit two possible branches of the square root
when solving (\ref{mu}) with respect to $\alpha _{q}$. For real
$\alpha _{q}$ it means the option of some negative $\alpha $'s
won't be ignored. We introduce
the set of signs of $\alpha _{q}$, $s_{q}=\{\pm 1\},q=2,...,k-1$ such that $%
s_{q}=s_{k+1-q}$. Then we get the solutions in the following general form%
\begin{eqnarray}
a_{q}(x) &=&\left( \prod\limits_{j=2}^{q}s_{j}\right) ^{q-1}\left( \frac{x}{%
\sqrt{\alpha ^{2}-1}}\right) ^{\sum_{j=1}^{q-1}r_{j}}\frac{\left(
x^{r_{q}}-1\right) }{\alpha },\;q=2,...,k-1,  \label{eq} \\
a_{k}(x) &=&\left( \prod\limits_{j=2}^{k-1}s_{j}\right) ^{k-2}\left( \frac{x%
}{\sqrt{\alpha ^{2}-1}}\right)
^{\sum_{j=1}^{k-2}r_{j}}\frac{x(x-1)}{\alpha ^{2}-1-x}  \label{ek}
\end{eqnarray}%
Note that an important artefact of the baxterization procedure in
the case of the Fuss-Catalan algebras is a necessary additional
requirement that all the parameters $\alpha _{q}$ of different
copies of TL subalgebras should be equal (up to some signs). Here
and below we denote $\alpha _{1}$ as $\alpha $ - the only free
parameter remaining after baxterization, which, we in principle
permit to be negative. This possibility of negative $\alpha $(s)
will extend variety of possible physical regimes of our
statistical models.

In \cite{DF} the solutions with $s_{q}=r_{q}=1$ were called
fundamental. All others differ from it either by less powers of
$x$ appearing in $a_{q}$, or alternating signs of $a_{q}$. Both
type of changes compared to the
fundamental solution should respect the symmetries of parameters $r$ and (\ref{alal}%
). The fundamental solution can be written as
\begin{equation}
W_{i}(x)=1_{i}+\sum_{q=1}^{k-1}\frac{x^{q-1}}{\left( \sqrt{\alpha ^{2}-1}%
\right) ^{q-1}}\frac{x-1}{\alpha }U_{i}^{(q)}+\frac{x^{k-1}}{\left( \sqrt{%
\alpha ^{2}-1}\right) ^{k-2}}\frac{x-1}{\alpha
^{2}-1-x}U_{i}^{(k)} \label{f}
\end{equation}

All the solutions for Boltzmann weights (\ref{w}) we got have two
important properties:
unitarity%
\begin{equation}
W_{i}(x)W_{i}(1/x)=1_{i}  \label{u}
\end{equation}%
and crossing symmetry:%
\begin{equation}
\overline{W}_{i}(x_{\ast }^{2}/x)=\left( \frac{x_{\ast
}}{x}\right) ^{\sum_{j=1}^{k-2}r_{j}}\frac{x_{\ast
}^{2}-x}{x(x-1)}W_{i}(x)  \label{cr}
\end{equation}%
which becomes
\begin{equation}
\overline{W}_{i}(x_{\ast }^{2}/x)=\left( \frac{x_{\ast }}{x}\right) ^{k-2}%
\frac{x_{\ast }^{2}-x}{x(x-1)}W_{i}(x)  \label{crf}
\end{equation}%
for the fundamental models. Here $x_{\ast }=$ $\sqrt{\alpha
^{2}-1}$, the
"bar" operation means $\overline{U}%
_{i}^{(q)}=U_{i}^{(k-q)},q=1,...,k,(U_{i}^{(0)}\equiv 1_{i})$. In
pictorial representation "bar" means rotation of TL generators by
90 degrees, as in
the crossing relation of Boltzmann weights of a lattice models. These two symmetries (\ref%
{u}),(\ref{cr}) are very important and play crucial role in
derivation of inversion relation. Here and below we mainly stop on
few representative
examples of the plenty solutions described by (\ref{eq}),(\ref{ek}),(\ref{f}%
). We will analyze in details $k=2,3,4$, considering possible
regimes of the models.

\subsection{Examples and regimes}

Here we write down explicitly all the solutions found for
$k=2,3,4$. Two color
case has only fundamental solution%
\begin{equation}
W_{i}^{(2)}(x)=1_{i}+\frac{x-1}{\alpha
}U_{i}^{(1)}+\frac{x(x-1)}{\alpha ^{2}-1-x}U_{i}^{(2)}  \label{k2}
\end{equation}%
In the case of three colors there are two possibilities%
\begin{equation}
W_{i}^{(3\pm )}(x)=1_{i}+\frac{x-1}{\alpha }U_{i}^{(1)}\pm \frac{x(x-1)}{%
\alpha \sqrt{\alpha ^{2}-1}}U_{i}^{(2)}\pm
\frac{x^{2}(x-1)}{\sqrt{\alpha ^{2}-1}\left( \alpha
^{2}-1-x\right) }U_{i}^{(3)}  \label{k3}
\end{equation}%
where the\ case of upper signs $(3+)$ corresponds to what we
called the fundamental solution. There are more possibilities in
the case of four colors. It is the minimal number of colors when
what we called an excited
solution (some $r_{q}=-1$) exists. Two solutions are%
\begin{eqnarray}
W_{i}^{(4\pm )}(x) &=&1_{i}+\frac{x-1}{\alpha }U_{i}^{(1)}\pm \frac{x(x-1)}{%
\alpha \sqrt{\alpha ^{2}-1}}U_{i}^{(2)}+\frac{x^{2}(x-1)}{\alpha
\left(
\alpha ^{2}-1\right) }U_{i}^{(3)}+  \nonumber \\
&&+\frac{x^{3}(x-1)}{\left( \alpha ^{2}-1\right) \left( \alpha
^{2}-1-x\right) }U_{i}^{(4)}  \label{k4f}
\end{eqnarray}%
The fundamental solution is $(4+)$ and $(4-)$ is an excited one.
There are more four color excited solutions:
\begin{eqnarray}
W_{i}^{(4^{\prime }\pm )}(x) &=&1_{i}+\frac{x-1}{\alpha
}U_{i}^{(1)}\pm
\frac{x-1}{\alpha }U_{i}^{(2)}+\frac{x-1}{\alpha }U_{i}^{(3)}+  \nonumber \\
&&+\frac{x(x-1)}{\left( \alpha ^{2}-1-x\right) }U_{i}^{(4)}
\label{k4e}
\end{eqnarray}

If we now impose positivity of all Boltzmann weights, we can find
different regimes of the models. It is convenient to give names to
the following regions on the plain $(\alpha ,x)$:
\begin{eqnarray}
I &:&\;1<x<\alpha ^{2}-1,\sqrt{2}<\alpha ,  \label{reg} \\
II &:&\;\alpha ^{2}-1<x<1,-\sqrt{2}<\alpha <-1,  \nonumber \\
III &:&\;0<x<1,-1<\alpha <0,  \nonumber \\
IV &:&\;-\infty <x<-1,-\infty <\alpha <-1,  \nonumber \\
V &:&\;-\infty <x<\alpha ^{2}-1,-1<\alpha <0,  \nonumber
\end{eqnarray}%
We will use the following parameterization of spectral and
crossing
parameters in the regions $I-V$:%
\begin{eqnarray}
I,II &:&\;x=e^{u},\;\alpha ^{2}-1=e^{\lambda },\;I: \ 0<u<\lambda,%
 \ II: \ \lambda <u<0. \label{co12} \\
III &:&\;x=e^{u},\;\alpha ^{2}-1=-e^{\lambda },\;\lambda <0,\;u<0
\label{co3} \\
IV &:&\;x=-e^{u},\;\alpha ^{2}-1=e^{\lambda },\;0<u  \label{co4} \\
V &:&\;x=-e^{u},\;\alpha ^{2}-1=-e^{\lambda },\;\lambda
<0,\;\lambda <u \label{co5}
\end{eqnarray}%
Models "$2$" and "$4^{\prime }+$" are physical (have positive
Boltzmann weights) in all the regimes $I-V$. "$3+$" is physical in
the regimes $I$ and $II$, and "$3-$" -- in the regimes $IV,V$.
"$4+$" has all Boltzmann weights positive only in the regime $I$,
whereas models "$4-$" and "$4^{\prime }-$" are never physical in
the sense of all Boltzmann weights positivity.

We see a rich spectrum of possibilities for different regions of
parameters defining a positive Boltzmann weights. Only detailed
investigation of the models can answer on the question what is the
phase structure of the models and its relation to the regimes.

\subsection{Vertex and IRF representations.}

In principle it would be plausible to proceed with exact solution
of these new integrable lattice models directly in terms of
Fuss-Catalan generators. Such possibility is shortly discussed in
the Sect.5. Here and below we formulate integrable lattice models
in terms of more concrete representations of the Fuss-Catalan
algebra. In this context the useful fact \cite{BJ} is that the
Fuss-Catalan algebra ( \ref{u1}),(\ref{u2}),(\ref{u3}) is a
subalgebra of tensor product of few copies of TL algebras with
different parameters $TL(\alpha _{1})\otimes ...\otimes TL(\alpha
_{k})$, and generators $U_{i}^{(q)}$ may be expressed
in terms of $u_{i}^{(q)}\in TL(\alpha _{q})$ as%
\begin{equation}
U_{i}^{(q)}=\left\{
\begin{array}{c}
1\otimes ...1\otimes u_{i}^{(k-q+1)}\otimes ...\otimes u_{i}^{(k)}
\ \ i \ - \ odd \\
u_{i}^{(1)}\otimes ...\otimes u_{i}^{(q)}\otimes 1\otimes
.....\otimes 1
\ \ i \ - \ even%
\end{array}%
\right.  \label{utp}
\end{equation}%
Recall that one of the requirements of the baxterization procedure
was the same value of $|\alpha _{q}|$ for different $q$.

The most studied lattice models are vertex models and interaction
round the face (IRF) models. In the first case the TL operator
$U_{i}^{(q)}$ acts on linear spaces attached to four links of
square lattice connected to each site. Using a basis $\{e_{i}\}$
of the linear space the vertex Boltzmann weights $W$ is defined as
\begin{equation}
W(x)e\otimes e=\sum_{i,j}W_{ij}^{kl}(x)e_{i}\otimes e_{j}
\label{vbw}
\end{equation}%
The fundamental representation for the usual TL generator $u$ is
known to be defined on a tensor product of two dimensional linear
spaces in a matrix
form:%
\begin{equation}
1=\left(
\begin{array}{cccc}
1 & 0 & 0 & 0 \\
0 & 1 & 0 & 0 \\
0 & 0 & 1 & 0 \\
0 & 0 & 0 & 1%
\end{array}%
\right) ,\;u=\left(
\begin{array}{cccc}
0 & 0 & 0 & 0 \\
0 & z & 1 & 0 \\
0 & 1 & 1/z & 0 \\
0 & 0 & 0 & 0%
\end{array}%
\right)  \label{tlv}
\end{equation}%
where $z$ \ is related to the TL algebra parameter $\alpha
=z+1/z$. In view of (\ref{utp}) it is natural to define
representation for the multicolored generator $U_{i}^{(q)}$ as the
tensor product of matrices (\ref{tlv}) (with,
in principle, different parameters $z$) and identity matrices, as in (\ref%
{utp}). The linear spaces living on the links of the lattice are
of dimension $2^{k}$, where $k$ is the number of colors.

Another representation of algebraically formulated lattice model Boltzmann weights $%
W_{i}(x) $ is an IRF representation. In this case some (integer)
numbers are living at each site of the square lattice with a
constraint that on two admissible (connected by a link) sites may
appear only numbers permitted by some fixed admissibility graph.
Then the representation is given (see e.g. \cite{Pa}) on an
elementary face of the lattice with site values $(i,j,k,l)$ in the
anti clockwise direction
\begin{equation}
1_{ij}^{lk}=\delta
_{ik},\;u_{ij}^{lk}=\frac{\sqrt{S_{i}S_{k}}}{S_{j}}\delta _{jl}
\label{tlf}
\end{equation}%
where $S_{i}$ are components of the eigenvector of the
admissibility matrix with the largest eigenvalue equal to the TL
parameter $\alpha $. $\alpha =2\cos \frac{\pi }{m},S_{i}=\sin
\frac{\pi i}{m},$ with $m$-Coxeter number for models with
$ADE$-like admissibility graphs, and $\alpha =2$ for the extended
$\widehat{A}\widehat{D}\widehat{E}$ admissibility graphs. Further
construction of representation of $U_{i}^{(q)}$ generators goes by
tensoring (\ref{tlf}) according to (\ref{utp}). This tensor
product of elementary IRF representations of TL operators will be
defined on a tensor product admissibility graph. As we said above,
not all admissibility graphs may participate in this tensor
product, since their matrix eigenvalues are not arbitrary, but
restricted by condition (\ref{alal}). Possible partners among
$ADE$ models which can be tensored, were classified in \cite{DF}.
Actually there is even more freedom -- one can make a tensor
product of vertex and IRF models, consistently fitting their
parameters $\alpha $. But in what follows we will concentrate on
two simplest examples -- two color vertex model and two color
$A_{n}$ IRF models.

The simplest possibility for vertex realization is the model $\
$"$2$". In has 32 non zero Boltzmann weights and list of them one
can find in \cite{DF}. Looking at this list, one can see that the
model is not of ice type -- there is no a natural "charge"
conservation in each vertex. Linear space living on each of four
edges of a vertex is four dimensional, such that R-matrix is
$16\times 16$. An interesting observation is that it is possible
to attach "charge" to these four spaces in a way preserving
"charge" conservation, but on a cost
of charge degeneracy of the spaces. For instance, the attachment: space $0$ $%
\rightarrow $ charge $0$, space $1$ $\rightarrow $ charge $1$, space $2$ $%
\rightarrow $ charge $1$, space $3$ $\rightarrow $ charge $3$ --
leads to "charge" conservation in each non zero vertex.

We formulate the model in the language of spin chain models, and
consider
the Boltzmann weights $W$ as an L-operator. (Recall that R-matrix is related to $L$%
-operator by permutation $P$ of outgoing spaces: $R=PL$, or $%
R_{ij}^{kl}=W_{ji}^{kl}$). We define the transfer matrix as%
\begin{equation}
T_{o(e)}(x)=\left( W_{o(e)}(x)\right)
_{c_{N}b_{1}}^{a_{1}c_{1}}\left( W_{e(o)}(x)\right)
_{c_{1}b_{2}}^{a_{2}c_{2}}\left( W_{o(e)}(x)\right)
_{c_{2}b_{3}}^{a_{3}c3}...\left( W_{e(o)}(x)\right)
_{c_{N-1}b_{N}}^{a_{N-1}c_{N}}  \label{tv}
\end{equation}%
There is a summation over repeated indices here, and indices $o/e$
mean
oddness(evenness) of the south site along the line of the auxiliary space $c$%
... As we see, $T$ \ is an alternating product of Boltzmann
weights on even and odd sites. (Recall that in the multicolored TL
models they are different). In the language of spin chains
Boltzmann weights may be considered as $L$-operators. Unitarity
condition looks like
\begin{equation}
\left( W_{o(e)}(x)\right) _{cd}^{ab}\left( W_{o(e)}(x^{-1})\right)
_{ef}^{cd}=\delta _{e}^{a}\delta _{f}^{b}  \label{ul}
\end{equation}%
There is an important property of BW, which is a mixture of crossing (\ref%
{cr}),(\ref{crf}) and unitarity relation:
\begin{equation}
W_{o(e)}(x)=\frac{x(x-1)}{x_{\ast }^{2}-x}M^{(1)}\left[
W_{e(o)}(x_{\ast }^{2}/x)\right] ^{t_{2}}M^{(1)}  \label{crl}
\end{equation}%
Here $t_{2}$ means transpose in the second space of Boltzmann
weights, corresponding to the auxiliary space of the transfer
matrix. The matrix $M^{(1)}$ acts in the first space and has the
form
\begin{equation}
M=\left(
\begin{array}{cccc}
0 & 0 & 0 & 1 \\
0 & 0 & 1/z & 0 \\
0 & 1/z & 0 & 0 \\
1/z^{2} & 0 & 0 & 0%
\end{array}%
\right)
\end{equation}%
Another important symmetry is%
\begin{equation}
\left( W_{o(e)}(x)\right) _{cd}^{ab}=PW_{e(o)}(x)P=\left(
W_{e(o)}(x)\right) _{dc}^{ba}  \label{p1}
\end{equation}%
The relations (\ref{ul}),(\ref{crl}) are almost enough for
derivation of the inversion relation (see below).

As the simplest example of two color IRF models one can consider
the case of tensor product admissibility graph $A_{n}\times
A_{n}$. The general
form of the RSOS Boltzmann weights are defined by (\ref{k2}),(\ref{utp}),(%
\ref{tlf})
\begin{eqnarray}
W_{e(o)}\left( ii^{\prime },jj^{\prime },kk^{\prime },ll^{\prime
}|x\right) &=&\delta _{ik}\delta _{i^{\prime }k^{\prime
}}+\frac{x-1}{\alpha }\left\{
\begin{tabular}{l}
$\delta _{i^{\prime }k^{\prime }}\delta
_{jl}\frac{\sqrt{S_{i}S_{k}}}{S_{j}}$
\ \ even site \\
$\delta _{ik}\delta _{j^{\prime }l^{\prime
}}\frac{\sqrt{S_{i^{\prime
}}S_{k^{\prime }}}}{S_{j^{\prime }}}$ \ \ odd site%
\end{tabular}%
\right. +  \label{irf} \\
&&+\frac{x(x-1)}{\alpha ^{2}-1-x}\delta _{jl}\delta _{j^{\prime }l^{\prime }}%
\frac{\sqrt{S_{i}S_{k}S_{i^{\prime }}S_{k^{\prime
}}}}{S_{j}S_{j^{\prime }}} \nonumber
\end{eqnarray}%
Pairs of variables $\left( i,i^{\prime }\right) ,\left(
j,j^{\prime }\right) ,\left( k,k^{\prime }\right) ,\left(
l,l^{\prime }\right) $, siting at the south, east, north and west
corners of a face, are any admissible set of
integers in the range from $1$ to $n$. Each of two quartets $(i,j,k,l)$ and $%
(i^{\prime },j^{\prime },k^{\prime },l^{\prime })$ are any permitted by $%
A_{n} $ admissibility graph diagram independently, and $S_{k}=\sin
\frac{\pi k}{n+1},\alpha =2\cos \frac{\pi }{n+1}$. One can
explicitly check that these
Boltzmann weights satisfy the following unitarity and crossing relations%
\begin{equation}
\sum_{k,k^{\prime }}W_{e(o)}\left( ii^{\prime },jj^{\prime
},kk^{\prime },ll^{\prime }|x\right) W_{o(e)}\left( kk^{\prime
},jj^{\prime },mm^{\prime },ll^{\prime }|1/x\right) =\delta
_{im}\delta _{i^{\prime }m^{\prime }} \label{irfu}
\end{equation}%
\begin{equation}
W_{e(o)}\left( ii^{\prime },jj^{\prime },kk^{\prime },ll^{\prime
}|x_{\ast
}^{2}/x\right) =\frac{\alpha ^{2}-1-x}{x(x-1)}\sqrt{\frac{%
S_{j}S_{l}S_{j^{\prime }}S_{l^{\prime }}}{S_{i}S_{k}S_{i^{\prime
}}S_{k^{\prime }}}}W_{o(e)}\left( jj^{\prime },kk^{\prime
},ll^{\prime },ii^{\prime }|x\right)  \label{irfcu}
\end{equation}

An interesting and important feature of the Boltzmann weights
(\ref{irf}) is their admissibility graph decomposition property.
One can check that admissibility rules coming from (\ref{irf})
lead to two independently existing subsets of site variables, each
with its own admissibility graph. The form of these two graphs
depends on oddness of $n$ (see Fig.1a, Fig.1b).

\begin{figure}[tbp]
\epsfysize=2.0 true in \hskip 25 true pt \epsfbox{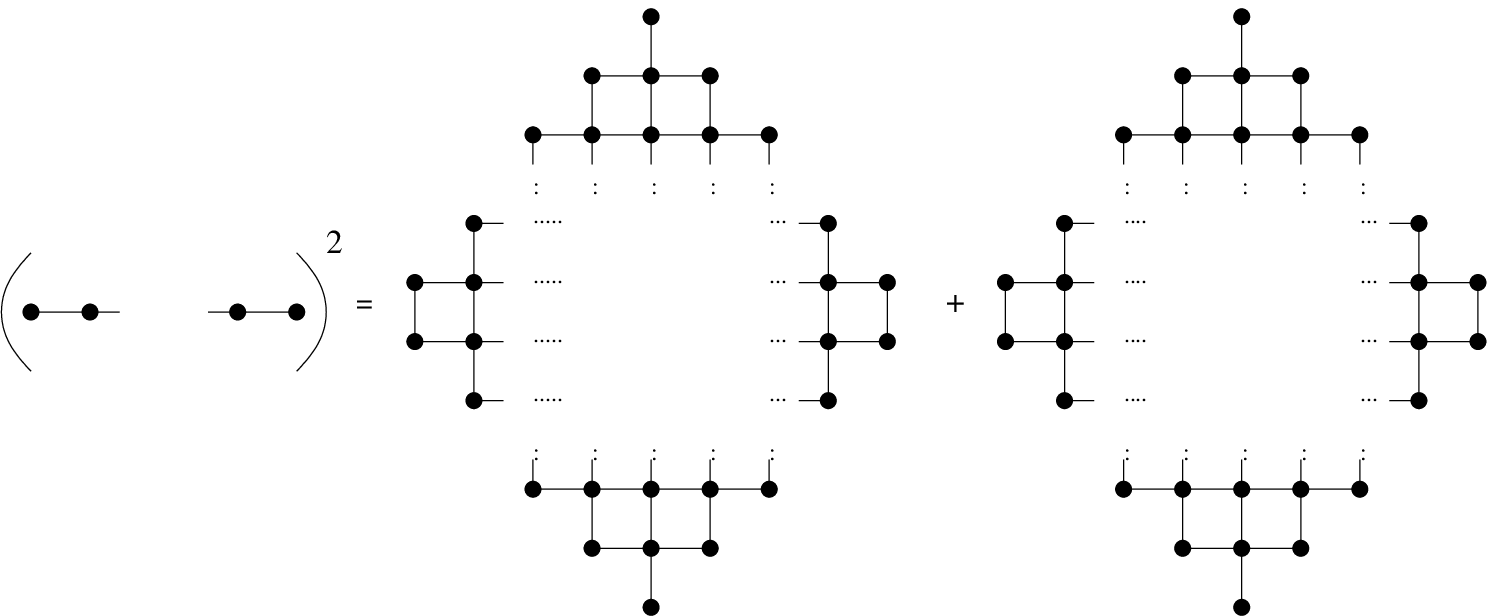}
\epsfysize=2.0 true in \hskip 25 true pt \epsfbox{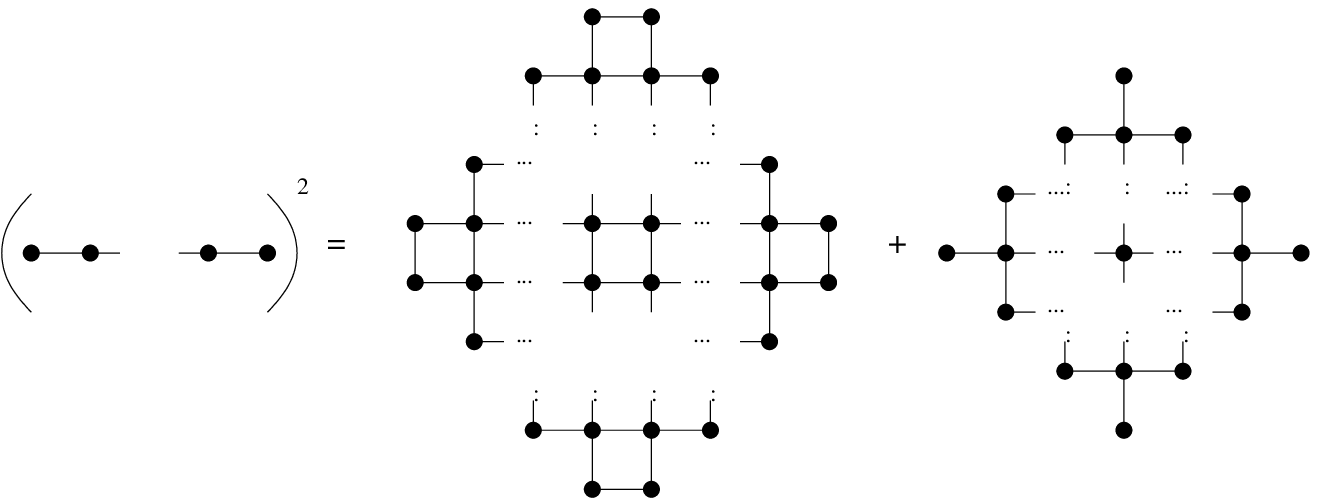}
\caption{$(A_n)^2$ admissibility graph decomposition:
\textbf{\ a} (upper) $n$ is even, \textbf{\ b} (lower) $n$ is odd}
\label{F1}
\end{figure}


Again, the basic object of investigation will be the transfer matrix%
\begin{eqnarray}
T_{o(e)}(x)_{\{\alpha \alpha ^{\prime }\}}^{\{\beta \beta ^{\prime
}\}} &=&\prod\limits_{i=1}^{N/2}W_{o(e)}(\alpha _{i}\alpha
_{i}^{\prime },\alpha _{i+1}\alpha _{i+1}^{\prime },\beta
_{i+1}\beta _{i+1}^{\prime },\beta
_{i}\beta _{i}^{\prime }|x)\times  \nonumber \\
&&W_{e(o)}(\alpha _{i+1}\alpha _{i+1}^{\prime },\alpha
_{i+2}\alpha _{i+2}^{\prime },\beta _{i+2}\beta _{i+2}^{\prime
},\beta _{i+1}\beta _{i+1}^{\prime }|x)  \label{tmirf}
\end{eqnarray}%
where the trace identification $\alpha _{0}=\alpha _{N},\alpha
_{0}^{\prime }=\alpha _{N}^{\prime },\beta _{0}=\beta _{N},\beta
_{0}^{\prime }=\beta _{N}^{\prime }$ is imposed. Recall here an
important and well known difference in definition of TM for vertex
and IRF cases. In the vertex case the auxiliary space of TM
contains a sum over thermodynamically many ($N$) variables $c_{i}$
in (\ref{tv}), whereas the IRF TM does not contain any auxiliary
space.

Let us look in more details at the two color $\left( A_{3}\right)
^{2}$ graph IRF model. This model, in a sense, may be called
$A_{3}$"two color" Ising model, since, as it is well known (see
e.g. \cite{DF1}), $A_{3}$ graph IRF model is equivalent to the
Ising model. Recall that $A_{3}$ admissibility rule requires that
the height value is fixed and equal to 2 (the value of the middle
site of $A_{3}$ graph) on the even sublattice of two dimensional
square lattice, whereas the heights on the odd sublattice take the
values 1 and 3, forming an Ising model. The situation is different
in the two color case and the model seems to be more complicated
then Ising
model. As one can see by explicit analysis of non zero Boltzmann weights for the case $n=3$%
, the RSOS model splits into two subsets of variables living on
the sites of a face. If one denotes ($\{\overline{\alpha }\}$) the
pairs of two numbers
at the sites of a face as%
\begin{equation}
(ij)\equiv \overline{\alpha }=3(i-1)+j  \label{coun}
\end{equation}%
the decomposition of the admissibility graph into two ones looks
like it is shown on the Fig.2. It means that there are either even
or odd heights in the new notation, living on each site of the
model. It means the model factorizes into two separate submodels.

\begin{figure}[tbp]
\epsfysize=2.0 true in \hskip 25 true pt \epsfbox{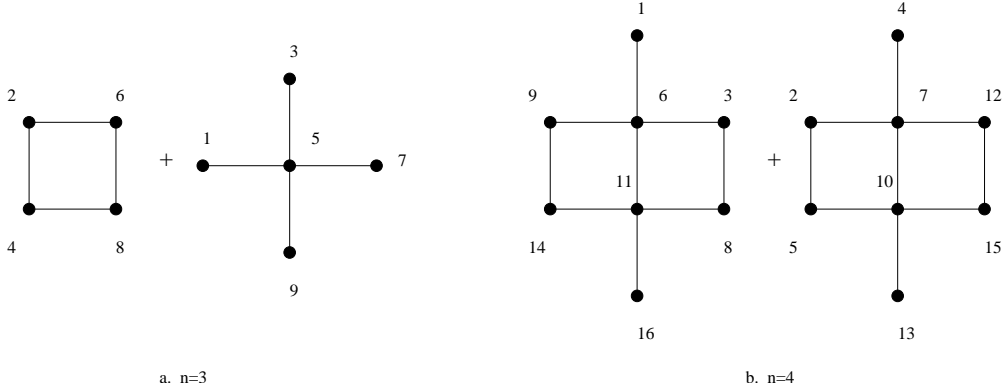}
\caption{Admissibility graph decomposition for $(A_3)^2$ and $(A_4)^2$}
\label{F2}
\end{figure}

In the same way, admissibility graph of the two color $\left(
A_{4}\right) ^{2}$ graph IRF model, after the recounting of the
site variables
\begin{equation}
(ij)\equiv \overline{\alpha }=4(i-1)+j
\end{equation}%
decomposes into two disconnected graphs (see Fig.4a,b), dividing
the model
into two submodels -- one with the site variables $\overline{\alpha }%
=\{1,3,6,8,9,11,14,16\}$, another -- with $\overline{\alpha }%
=\{2,4,5,7,10,12,13,15\}$. Again there are two separate submodels.

Note that the relations (\ref{irfcu}) are valid for each of two
submodels of $(A_{n})^{2}$ admissibility graph decomposition.

\section{Inversion functional relations and their solutions.}

The most full and contemporary investigation of the models
described in the previous section should probably follow the
standard methods of algebraic Bethe ansatz. Unfortunately the
obstructions we met on this way didn't allow us to proceed in this
direction, and algebraic Bethe ansatz for these models remains an
open problem. But some information about ground state energy and
free energy, spectrum of low lying excitations, and even about
some correlation functions, can be extracted using a method known
for a long time \cite{B1}, which doesn't require a knowledge of
explicit form BA equations and transfer matrix eigenvalues. It is
a method of inversion functional relations on the transfer matrix
eigenvalues, which was successfully applied to few integrable
lattice models \cite{K}. It would be interesting to develop a
universal algebraic method of derivation of inversion relations in
terms of multicolored TL algebra generators, as it was partly done
in \cite{PM} for usual TL algebra models, but in what follows we
are going to consider the inversion relation in vertex and IRF
representations of multicolored TL algebra, which we discussed in
the previous section.

\subsection{Vertex model}

We start from the vertex model. Consider the product
\begin{equation}
\left( M\right) ^{\otimes N}T_{o}(x_{\ast }^{2}x)\left( M\right)
^{\otimes N}T_{e}(x)
\end{equation}%
Since transfer matrices commute, their eigenvalues will satisfy
the same functional relations as transfer matrices themselves. The
above expression explicitly looks as
\begin{eqnarray*}
&&Tr_{A}Tr_{B}\left[ M^{C_{1}}W_{oAC_{1}}^{t_{A}}(x_{\ast
}^{2}x)M^{C_{1}}M^{C_{2}}W_{eAC_{2}}^{t_{A}}(x_{\ast
}^{2}x)M^{C_{2}}...\ast
\right. \\
&&\left. W_{eBC_{1}}(x)W_{oBC_{2}}(x)...\right]
\end{eqnarray*}%
Insertion of the equal to 1 factor $(\delta _{a_{i}b_{i}}+1-\delta
_{a_{i}b_{i}})$ in any place of this product leads to a split of
the above expression into two terms -- the first corresponding to
the $\delta _{a_{i}b_{i}}$, and the second -- to $\left( 1-\delta
_{a_{i}b_{i}}\right) $.
 Due to the relations (\ref{crl}),(\ref{ul}) the first term turns
out to be proportional to the identity operator in all the $C_{i}$
spaces. Oppositely, the second term is not giving an identity
operator in any of $C_{i}$ spaces. According to (\ref{crl}), the
proportionality coefficient for the first term
will be $\frac{1}{z^{2N}}\left( \frac{x_{\ast }^{2}-x^{-1}}{x^{-1}(x^{-1}-1)}%
\right) ^{N}$. One can show that the expansion of the second term
in powers of $x$ starts from the $x^{N/2}$, therefore in the
region where $x>1$, e.g. in the regime $I$, the second term is
exponentially small compared to the first term in the
thermodynamic limit. This leads to the following inversion
relation
\begin{equation}
\left( M\right) ^{\otimes N}T_{o}(x_{\ast }^{2}x)\left( M\right)
^{\otimes
N}T_{e}(x)=\frac{1}{z^{2N}}\left( \frac{x_{\ast }^{2}-x^{-1}}{%
x^{-1}(x^{-1}-1)}\right) ^{N}I_{C_{1}}\otimes I_{C_{2}}\otimes
.....I_{C_{N}}+O(e^{-N})
\end{equation}%
One can also see that specific of the matrix $M$ is such that the
relation between highest eigenvalue of $\left( M\right) ^{\otimes
N}T_{o}(x_{\ast }^{2}x)\left( M\right) ^{\otimes N}$ and of
$T_{o}(x_{\ast }^{2}x)$ is by factor $\frac{1}{z^{2N}}$. It leads
to the following functional relation for eigenvalues of transfer
matrix in the thermodynamic limit $N\rightarrow \infty $, when the
terms $O(e^{-N})$ can be neglected
\begin{equation}
\Lambda (x_{\ast }^{2}x)\Lambda (x)=\left( \frac{x_{\ast }^{2}-x^{-1}}{%
x^{-1}(x^{-1}-1)}\right) ^{N}  \label{inv1}
\end{equation}

Another relation is the unitarity relation. If one considers the
product
\begin{equation}
T_{o}(x)T_{e}(1/x)=Tr_{A}Tr_{B}[W_{oAC_{1}}(x)W_{eAC_{2}}(x)...W_{eBC_{1}}(1/x)W_{oBC_{2}}(1/x)...]
\end{equation}%
then by insertion of the space permutation operator $P$
($P^{2}=1$) between the Boltzmann weights of, say, the second TM,
and using (\ref{p1}), one gets
\begin{equation}
\sum_{\{a\},\{b\},\{c%
\}}W_{o}(x)_{a_{N}c_{1}}^{d_{1}a_{1}}W_{e}(x)_{a_{1}c_{2}}^{d_{2}a_{2}}...W_{o}(1/x)_{e_{1}b_{N}}^{b_{1}c_{1}}W_{e}(1/x)_{e_{2}b_{1}}^{b_{2}c_{2}}...
\end{equation}

If one inserts one delta symbol $\delta _{e_{i+1}a_{i}}$ for some
$i$, then due to the relation (\ref{ul}) all the product will be
reduces to the identity operator $Id=\delta _{e_{1}d_{1}}...\delta
_{e_{N}d_{N}}$. One can also show that other terms successfully
cancel. This leads to the second functional relation on the
transfer matrix eigenvalues
\begin{equation}
\Lambda (x)\Lambda (1/x)=1  \label{inv2}
\end{equation}

The functional equations (\ref{inv1}),(\ref{inv2}) can be solved
by standard methods. These equations in the parameterization
(\ref{co12}) for regime $I$, can be written in terms of
$\overline{\Lambda }(u)=\Lambda ^{1/N}(x)$, as
\begin{eqnarray}
\overline{\Lambda }(u)\overline{\Lambda }(\lambda +u)
&=&\frac{e^{\lambda
+u}-1}{e^{-u}-1}  \label{sys1} \\
\overline{\Lambda }(u)\overline{\Lambda }(-u) &=&1  \nonumber
\end{eqnarray}

In order to solve uniquely this system of functional equations,
one needs
additional information about analyticity of the function $\overline{\Lambda }%
(u)$. One can look for solutions in the form $\overline{\Lambda }%
(u)=F(u)f(u) $, where $F(u)$ is a general analytic non zero (ANZ)
in the quadrant $0<Re(u)<\lambda ,0<Im(u)<2\pi $ solution of the
system
\begin{eqnarray}
F(u)F(\lambda +u) &=&\frac{e^{\lambda +u}-1}{e^{-u}-1}  \label{sys} \\
F(u)F(-u) &=&1  \nonumber
\end{eqnarray}%
and $f(u)$ is a solution of%
\begin{eqnarray}
f(u)f(u+\lambda ) &=&1  \label{sys2} \\
f(u)f(-u) &=&1  \nonumber
\end{eqnarray}%
with a given set of poles $u_{p}$ and zeros $u_{z}$ in the same
quadrant. We have cut the strip $0<Im(u)<2\pi$ in the imaginary
direction because of the obvious $2\pi i$ periodicity of solutions
of the system (\ref{sys1}). The cut in the real axis direction
$0<Re(u)<\lambda $ is a consequence of the relation $f(u+2\lambda
)=f(u)$, which one can easily derive from (\ref{sys2}).

The general ANZ solution of (\ref{sys}) has the form
\begin{equation}
F(u)=\left( e^{u}-1\right) \prod\limits_{j=0}^{\infty
}\frac{\left( e^{(2j+1)\lambda +u}-1\right) \left(
e^{(2j+2)\lambda -u}-1\right) }{\left( e^{2j\lambda +u}-1\right)
\left( e^{(2j+1)\lambda -u}-1\right) }  \label{s1}
\end{equation}

For the double periodic function $f(u)$ we have the system of
functional equations (\ref{sys2}). Provided we know the set of its
poles and zeros in the periodicity quadrant, the solution is
uniquely fixed by Liouville's
theorem:%
\begin{equation}
f(u)=\pm \prod\limits_{u_{z}}\sqrt{\kappa _{1}}snh\left( \frac{2K_{1}}{\pi }%
(u-u_{z})\right) \prod\limits_{u_{p}}\sqrt{\kappa _{1}}snh\left( \frac{%
2K_{1}}{\pi }(u-u_{p}-\lambda )\right)  \label{zp}
\end{equation}%
where $snh$ is a standard elliptic function with modulus $\kappa
_{1}$
defined by the requirement that corresponding quarter periods $%
K_{1},K_{1}^{\prime }$ are related by $\frac{K_{1}^{\prime }}{K_{1}}=\frac{%
\lambda }{\pi }$. (\ref{s1}) and (\ref{zp}) give general form of
transfer matrix highest eigenvalue as a function of spectral
parameter $u$, provided one knows positions of its poles and zeros
in the periodicity region. Obviously transfer matrix eigenvalues
do not have poles in the physical region. Some numerics we did for
the lattices of small size in regime $I$ support the conjecture
that the ground state eigenvalue do not have also zeros in the
periodicity region, and $f(u)=1$ for the ground state. Using
(\ref{s1}) one can calculate the
ground state energy of the model%
\begin{equation}
E_{0}=-\frac{d}{du}\left( \ln \overline{\Lambda }(u)\right)
_{u=0}=-\sum_{j=0}^{\infty }\frac{\sinh \frac{\lambda }{2}}{\sinh
(j+1)\lambda \sinh (j+\frac{1}{2})\lambda }  \label{e0}
\end{equation}

\subsection{IRF model}


Discussing the family of IRF models, we will concentrate on two color $%
(A_{n})^{2}$ models. In the same way, the product of two transfer
matrices (\ref{tmirf}) for the IRF model has the form
\begin{eqnarray*}
T_{o}(x_{\ast }^{2}x)T_{e}(x) &=&\sum_{\{\overline{\alpha }%
_{i}\}}\prod\limits_{i=1}^{N/2}W_{o}(\overline{\alpha }_{i},\overline{%
\alpha }_{i+1},\overline{\beta }_{i+1},\overline{\beta
}_{i}|x_{\ast
}^{2}x)W_{e}(\overline{\alpha }_{i+1},\overline{\alpha }_{i+2},\overline{%
\beta }_{i+2},\overline{\beta }_{i+1}|x_{\ast }^{2}x)\ast \\
&&W_{e}(\overline{\gamma }_{i},\overline{\gamma }_{i+1},\overline{\alpha }%
_{i+1},\overline{\alpha }_{i}|x)W_{o}(\overline{\gamma }_{i+1},\overline{%
\gamma }_{i+2},\overline{\alpha }_{i+2},\overline{\alpha
}_{i+1}|x)
\end{eqnarray*}%
Using crossing and unitarity properties (\ref{irfcu}) of the BW
one can
easily see that insertion of one Kronecker $\delta _{\overline{\alpha }_{i}%
\overline{\gamma }_{i}}$ for some one $i$ into the last
expression, reduces
it to the identity operator for all $i$ with the prefactor $\left( \frac{%
x_{\ast }^{2}-x^{-1}}{x^{-1}(x^{-1}-1)}\right) ^{N}$. It means
that the identity element is in a sense "decoupled" from the
others. With some effort, one can show that, like in the vertex
case, $x$ expansion of operators different from identity starts
from $x$ in the maximal power less then $N$. It leads to the same
functional relation as (\ref{inv1}).
\begin{equation}
T_{o}(x_{\ast }^{2}x)T_{e}(x)=\left( \frac{x_{\ast }^{2}-x^{-1}}{%
x^{-1}(x^{-1}-1)}\right) ^{N}+O(e^{-N})  \label{inv3}
\end{equation}%
Unfortunately in the IRF case the situation is different with the
unitarity relation (\ref{inv2}). Were we try to generate an
identity operator for product of two transfer matrices
$T_o(x)T_e(1/x)$, we would need to introduce $2N$ delta symbols in
order to make the relation (\ref{irfu}) working, without involving
the crossing relation (\ref{irfcu}). (Recall that there is no any
auxiliary space trace in the definition of the IRF transfer
matrix). Apparently such drastic deformation operator cannot be
negligibly small in the thermodynamic limit $N\rightarrow \infty$.
There is the well known way to avoid this obstacle: one can pass
to the bigger object -- partition function.

\begin{figure}[tbp]
\epsfysize=3.0 true in \hskip 25 true pt \epsfbox{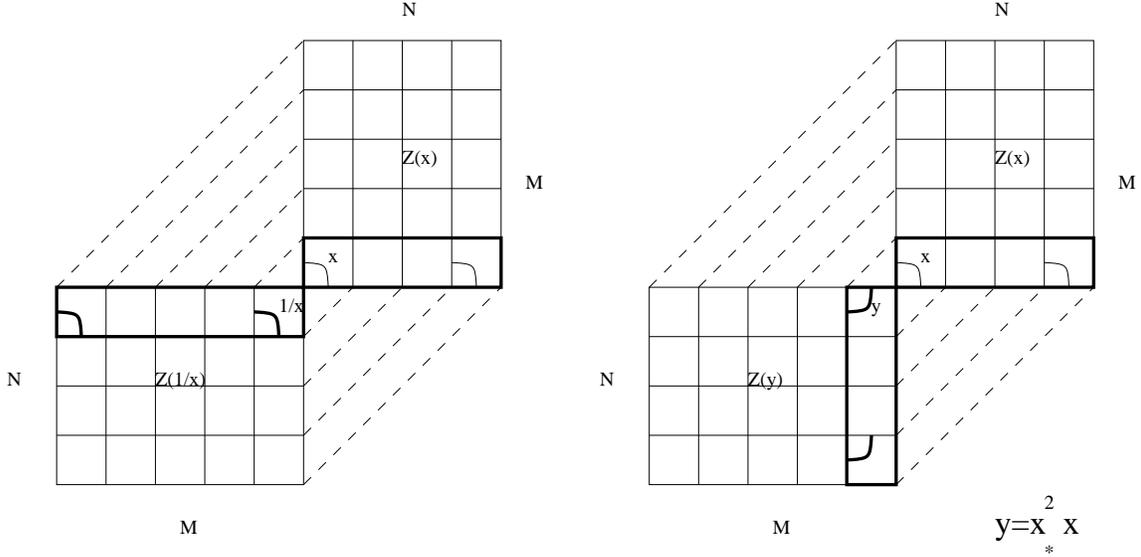}
\caption{Illustration of inversion relation derivation for IRF
model} \label{F3}
\end{figure}

Consider partition function for $2N\times 2M$ square lattice. It
can be written in different ways
\[
Z(x)=\sum_{p.b.c.}\left(T_o^{(N)}(x)
T_e^{(N)}(x)\right)^{2M}=\sum_{p.b.c.}\left(T_o^{(M)}(x)
T_e^{(M)}(x)\right)^{2N}
\]
where the transfer matrix product means summation over the
repeating site variables of the multiplied transfer matrices, and
$\sum_{p.b.c.}$ summation over periodic boundary conditions in
both directions. One can introduce a set $g_{M+N}$ of delta
symbols matching the boundary configurations for both statistical
sums (see the dashed lines on the illustrating Fig.3 left).
Insertion of the identity operator in the form
$1=g_{M+N}+(1-g_{M+N})$ separates the partition functions product
into two pieces:
\begin{equation}
Z(1/x) Z(x)=\sum \left(T_o^{(M)}(1/x) T_e^{(M)}(1/x)\right)^{2N}
g_{M+N} \left(T_o^{(N)}(x) T_e^{(N)}(x)\right)^{2M}+O(x)
\label{uin}
\end{equation}
One can see that the first term "collapses" into operator
identifying configurations on all the sites of both lattices due
to the relation (\ref{irfu}). This "collapse" goes plaquette by
plaquette, without any relation to the transfer matrices. The
second term $O(x)$ represents deviations from identity. Here we
will \textit{assume} that the boundary changing operator $g_{M+N}$
produces only subdominant terms for partition function. This
assumption was shown to be correct for some IRF models \cite{K},
but investigation of this question is needed in the case of our
IRF models. With this assumption the first term in (\ref{uin}) is
dominating over $O(x)$.

In the similar way one can translate the relation (\ref{inv3})
valid by itself, to the language of the partition functions
product. One can see (for illustration see the Fig.3 right) that
the same $g_{M+N}$ does the job:
\begin{equation}
Z(x_{*}^2 x) Z(x)=\sum \left(T_o^{(N)}(x_{*}^2 x)
T_e^{(N)}(x_{*}^2 x)\right)^{2M} g_{M+N} \left(T_o^{(N)}(x)
T_e^{(N)}(x)\right)^{2M}+O(x) \label{ucin}
\end{equation}
In the first term, as it is illustrated on the Fig.3b, the product
of the right column transfer matrix of $Z(x_{*}^2 x)$ with the
lowest row transfer matrix of $Z(x)$ by (\ref{inv3}) is
proportional to the identity operator. This "collapse" to the
identity continues to the product of the next to right column with
the next to lowest row, and so on. By the same reasons as in the
previous (\ref{uin}) relation, one can believe that the first term
is dominating in the thermodynamic limit.

The dominating terms of equations (\ref{uin}),(\ref{ucin}) lead to
the system of functional equations for partition function density
$z(x)=Z(x)^{1/MN}$
\begin{eqnarray}
z(x)z(1/x)&=&1 \label{irfsys} \\
z(x_{*}^2 x) z(z)&=&\frac{x_*^2-x^{-1}}{x^{-1}(x^{-1}-1)}
\nonumber
\end{eqnarray}
which coincides with the equations (\ref{sys1}) in the proper
regime parameterization. Numerical simulations show that for two
color $(A_{3})^{2}$ and $(A_{4})^{2}$ models $z(u)$ is ANZ in the
periodicity domain. It means that the solution for partition
function in the ground state $z(u)$ is given by (\ref{s1}). An
interesting property of it is that $z(0)=1$.

\subsection{Spectrum of low lying excitations}

In the same way one can consider the low lying excitations, i.e.
those states for which the transfer matrix eigenvalue $\Lambda
(u)$ is finitely different from the ground state one $\Lambda
_{gs}(u)$ in the thermodynamic limit. For instance, define finite
function
\begin{equation}
l(u)=\lim_{N\rightarrow \infty }\frac{\Lambda (u)}{\Lambda
_{gs}(u)}
\end{equation}%
for vertex model. Again $l(u)$ can be represented as a product of
ANZ function $F(u)$ and another one $f(u)$ satisfying the same
functional equations (\ref{sys}),(\ref{sys2}). But this time
$f(u)\neq 1$, and some zero(s) $u_{z}$ enter into the periodicity
domain. The excitation spectrum can be calculated from the
relation between energy and momentum of excitations
\begin{eqnarray*}
E-E_{0} &=&-\sum_{\{u_{z}\}}\frac{d}{du}(\ln \psi
(u))_{u=0}=\sum_{\{u_{z}\}}\varepsilon (p(u_{z})) \\
P-P_{0} &=&-i\sum_{\{u_{z}\}}\ln \psi (0)=\sum_{\{u_{z}\}}p(u_{z})
\end{eqnarray*}%
where $\ln \psi (u)$ is the contribution of each $snh$ factor in
the product
(\ref{zp}). Using (\ref{zp}) we get the dispersion relation:%
\begin{equation}
\varepsilon (p)=K^{\prime }\sqrt{(1-\kappa )^{2}+4\kappa \sin
^{2}p} \label{dis}
\end{equation}%
with non zero mass gap%
\begin{equation}
\Delta =K^{\prime }(1-\kappa )  \label{gap}
\end{equation}%
So, at least in the regime $I$, where one can neglect the
additional terms in the inversion relation, all the excitations
are with a non zero energy gap. It means there is no conformal
critical point in this regime.

Importance of the regime in all what was done above, is crucial in
two aspects. First, in order to neglect the non identity terms in
the inversion relation one needs $|x|>1$, which is correct not
only for regime $I$, but also for regime $IV$ and $x<-1$ subregion
of the regime $V$ (we will call it $V^{\prime }$). Second, the
numerics, which is necessary to be sure in ANZ
condition, essentially differ for different regimes. The analog of (\ref%
{sys1}) for vertex model in a parameterization
(\ref{co4}),(\ref{co5}) of the
regimes $IV,V^{\prime }$ looks like%
\begin{eqnarray}
\overline{\Lambda }(u)\overline{\Lambda }(\lambda +u)
&=&\frac{e^{\lambda
+u}-p}{e^{-u}+1}  \label{sys3} \\
\overline{\Lambda }(u)\overline{\Lambda }(-u) &=&1  \nonumber
\end{eqnarray}%
where in the regime $IV$ $p=-1$, and in the regime $V^{\prime }$
$p=1$. Equation (\ref{sys2}) remains as it is. It means the
solution for $F$ in this case is given by
\begin{equation}
F(u)=\left( e^{u}+1\right) \prod\limits_{j=0}^{\infty
}\frac{\left( e^{(2j+1)\lambda +u}-p\right) \left(
e^{(2j+2)\lambda -u}+1\right) }{\left( e^{2j\lambda +u}+1\right)
\left( e^{(2j+1)\lambda -u}-p\right) }  \label{so}
\end{equation}%
$f(u)$ is given by the same solution (\ref{zp}). With the
assumption of ANZ condition for $\Lambda (u)$ it leads to the
following ground state energies of the two color vertex model in
the regimes $IV$ and $V^{\prime }$:
\begin{equation}
E_{0}^{IV}=-\sum_{j=0}^{\infty }\frac{\cosh \frac{\lambda
}{2}}{\cosh (j+1)\lambda \cosh (j+\frac{1}{2})\lambda
},\;E_{0}^{V^{\prime }}=-\sum_{j=0}^{\infty }\frac{\cosh
\frac{\lambda }{2}}{\cosh (j+1)\lambda \sinh
(j+\frac{1}{2})\lambda }
\end{equation}%
 The analysis of low lying
excitations remains the same, giving the same finite energy gap
(\ref{gap}).

In the same way one can modify the solutions for partition
function $z(u)$ of two color IRF models in regimes $IV$ and
$V^{\prime }$. With an ANZ hypothesis, it is given by the same
expression (\ref{so}) with the same property $z(0)=1$.

\section{Other models.}

A natural conjecture is that all what was done in the previous
section for two colored \ models, can be generalized for what was
called above fundamental models with any number of colors in the
regime analogous to regime $I$.  Lets point out that both three
and four colored models have only regimes $I$ and $II$ as regimes
with all Boltzmann weights positive. As we said, $x>1$ only in the
regime $I$, which was crucial for derivation of inversion
relation. Instead of (\ref{crl}) we have
\begin{equation}
L_{i}(x)=\left( \frac{x_{\ast }}{x}\right)
^{k-2}\frac{x(x-1)}{x_{\ast
}^{2}-x}\widetilde{M}^{(1)}\left[ L_{i+1}(x_{\ast }^{2}/x)\right] ^{t_{2}}%
\widetilde{M}^{(1)}
\end{equation}%
with some matrix $\widetilde{M}$ acting in the first space.
 Together with (\ref{ul}) it gives the following inversion relations%
\begin{eqnarray*}
T(x)T(x_{\ast }^{2}x) &=&(x_{\ast }x)^{k-2}\left[ \frac{x_{\ast }-x^{-1}}{%
x^{-1}(x^{-1}-1)}\right] ^{N}Id+O(e^{-N}) \\
T(x)T(1/x) &=&Id
\end{eqnarray*}%
As we see, they differ from (\ref{inv3}) is minimal way -- by a
prefactor. The same procedure as previously gives
\begin{eqnarray*}
\Lambda (u) &=&\overline{\Lambda }(u)f(u) \\
\overline{\Lambda }(u) &=&e^{(k-2)u}\left( e^{u}-1\right)
\prod\limits_{j=0}^{\infty }\frac{\left( e^{(2j+1)\lambda
+u}-1\right) \left( e^{(2j+2)\lambda -u}-1\right) }{\left(
e^{2j\lambda +u}-1\right) \left( e^{(2j+1)\lambda -u}-1\right) }
\end{eqnarray*}%
with the same expression (\ref{zp}) for $f(u)$. It leads to
integer shift of the ground state energy in $k-2$, compared to
(\ref{e0}), provided still $f=1 $  (ANZ hypothesis). This of
course should be checked numerically.

Looking at the other models which have some physical regimes, one
can believe that the general form of inversion relations both for
fundamental models with some negative signs and for excited models
with or without negative signs, is just
\begin{eqnarray*}
T(x)T(x_{\ast }^{2}x) &=&(x_{\ast }x)^{\sum r_{j}-2}\left[
\frac{x_{\ast
}-x^{-1}}{x^{-1}(x^{-1}-1)}\right] ^{N}Id+O(e^{-N}) \\
T(x)T(1/x) &=&Id
\end{eqnarray*}%
Form of solution of these equations depends on regime. We see that
different models have the same set of functional inversion
relations, and sometimes even the same set of regimes.

One can believe that if there is a difference in ground state or
low lying excited states between these models, then it is
expressed in the amount and location of poles and zeros of
function $f(u)$. An information about these poles and zeros may
come from numerics.

\section{Discussion}

In this paper we made an attempt to analyze vertex and IRF
multicolored lattice models. We saw an interesting structure of
admissibility graph for IRF models. Being unable to solve the
models by algebraic Bethe ansatz, we tried to analyze them using
the inversion relations. The main result of this attempt - all the
models have a mass gap in the considered regime. Another evidence
for absence of conformal points in this regime we saw formally
trying to construct new link invariants, following a standard
procedure of extracting a braid group representation from the
Boltzmann weights either in vertex or in IRF form (see \cite{DWA}
and references therein). The objection we met in this way is that
it is impossible to get a well defined braid group representation
taking a limit $u\rightarrow i\infty $, or some other one, for the
two color Boltzmann weights $W(u)$. Also the formal sufficient
condition for construction of link invariants from $W(u)$ one can
find in \cite{DWA} is not satisfied. But one cannot reject a
possibility of gapless behavior for the models we considered in
other regimes, as it was pointed out in \cite{MN1} for dimerised
coupled spin chain.

Let us stress an almost unrestricted amount of possibilities in
construction of new integrable multicolored TL models demonstrated
above. Certainly these zoo of new models deserves a classification
which, as far as we know, is missing for today.

An interesting possible alternative is coordinate Bethe ansatz
formulated in terms of the Fuss-Catalan algebra itself. Such kind
of Bethe ansatz formulation, purely in terms of usual TL algebra
generators for the XXZ model, was applied in \cite{LMS}. The basis
for wave function representation was chosen as a basis of
specially defined two sided ideals of words of TL generators. It
would be interesting to apply such formulation to the multicolored
TL lattice models.

 This paper is only a first touch to these new
integrable models. Development \ and applications of known methods
of integrable models, such as algebraical Bethe ansatz, transfer
matrix functional relations leading to a (system of) integral
equations, a role of Q-operator in these models, and so on -- are
open and interesting problems.

\section{Acknowledgements}

A.B. is grateful to M.J.Martins for very useful communications.
We are thankful to Einstein Center for financial support.

\end{document}